\documentclass[twocolumn]{revtex4}
\usepackage{xcolor}
\usepackage{graphics,graphicx,epsfig}
\usepackage{epsf,epstopdf,wrapfig}
\usepackage{amssymb,amsfonts,amsmath}
\usepackage[export]{adjustbox}
\include{epsf}
\usepackage{ifthen}

\graphicspath{{./figures/},{./si_figures/}}

\newcommand{\beqn}{\begin{eqnarray}}
\newcommand{\eeqn}{\end{eqnarray}}
\newcommand{\beq}{\begin{equation}}
\newcommand{\eeq}{\end{equation}}

\newcommand{\<}{\langle}
\renewcommand{\>}{\rangle}

\usepackage{ifthen}

\begin{document}
\title{Population variability in the generation and thymic selection of T-cell repertoires}

\author{ Zachary Sethna$^{1}$, Giulio Isacchini$^{2,3}$, Thomas Dupic$^{2}$, Thierry Mora$^{2}$, Aleksandra M. Walczak$^{2}$, Yuval Elhanati$^{4}$}

\affiliation{
	\normalsize{$^{1}$Sloan Kettering Institute, Memorial Sloan Kettering Cancer Center, New York, NY, USA}\\
	\normalsize{$^{2}$ Laboratoire de physique de l'\'Ecole Normale Sup\'erieure, PSL University, CNRS,
          Sorbonne Universit\'e, Universit\'e de Paris,
          24 rue Lhomond, 75005 Paris, France}\\
        \normalsize{$^{3}$ Max Planck Institute for Dynamics and Self-organization, Am Fa\ss berg 17, 37077 G\"ottingen, Germany}\\
        \normalsize{$^{4}$Immunogenomics and Precision Oncology Platform, Human Oncology and Pathogenesis Program, Memorial Sloan Kettering Cancer Center, New York, NY, USA}\\
	}
	
\date{\today}

\begin{abstract}
The diversity of T-cell receptor (TCR) repertoires is achieved by a combination of two intrinsically stochastic steps: random receptor generation by VDJ recombination, and selection based on the recognition of random self-peptides presented on the major histocompatibility complex. These processes lead to a large receptor variability within and between individuals. However, the characterization of the variability is hampered by the limited size of the sampled repertoires. 
We introduce a new software tool SONIA to facilitate inference of individual-specific computational models for the generation and selection of the TCR beta chain (TRB) from sequenced repertoires of 651 individuals, separating and quantifying the variability of the two processes of generation and selection in the population. We find not only that most of the variability is driven by the VDJ generation process, but there is a large degree of consistency between individuals with the inter-individual variance of repertoires being about $\sim$2\% of the intra-individual variance. Known viral-specific TCRs follow the same generation and selection statistics as all TCRs.

\end{abstract}

\maketitle

\section{Introduction}

Most organisms live in a similar environment, facing common pathogenic threats. However, the adaptive immune system, based on the stochastic VDJ recombination process, is a naturally diverse system, supporting both repertoire variability within the individual, and variability across the population \cite{janeway2001principles}. Quantifying both types of variability, and understanding how they support a robust immune response, are still open questions. Determining the variability under normal healthy conditions is a crucial step for understanding the immune system in compromised situations such as infections, autoimmune diseases, and cancer. 

The adaptive immune system reacts specifically against a variety of different threats to the organism. This is achieved by maintaining a large ensemble of T cells, each having a different receptor that binds distinct subsets of antigens. The adaptive immune system maintains this diversity by generating a large repertoire of cells with different receptors \cite{Hozumi1976,Venturi2006,Murugan2012} and then selecting them according to their binding properties. The first step of selection occurs in the thymus. Cells carrying receptors that bind too strongly or too weakly to the host's own proteins do not pass this selection \cite{Kyewski2006, Starr2003}. The remaining cells are let out into the periphery and undergo selection for binding of foreign antigen which results in cell proliferation. In all cases, T cell receptors (TCR) bind to antigen fragments presented as short peptides on the major histocomptability complex (MHC) of presenting cells \cite{Clambey2014}. Each human individual has 6 types of MHC molecules encoded by the very polymorphic human leukocyte antigen (HLA) locus. All of these processes ---\,receptor generation, selection, and peptide presentation\,--- are stochastic in nature and depend on the host's genetic background.

High-throughput T cell repertoire sequencing (RepSeq) provides a census of the T cell repertoire found in a blood or tissue sample \cite{Heather2017,Lindau2017,Six2013,Woodsworth2013}. These samples are generally indicative of the true repertoire, and comparing them over a population yields similarities predicted based on MHCs, pathogenic history and general properties of the generation process \cite{DeWitt2018,Elhanati2018}. Due to the large diversity of possible TCRs, different samples, even ones taken from the same individual under the same conditions, will often differ substantially due to statistical noise. As a result, characterization of a repertoire sample is often more reliably done by statistically modeling the underlying generation and selection processes instead of working with raw TCR sequences and read counts. In this paper we take such an approach to characterize the diversity of the human T-cell receptor beta chain (TRB) repertoire. This approach allows us to disentangle the two processes of generation and selection, and to quantify their relative contribution to the overall variability across individuals. Our results provide a quantification of natural TCR diversity which is essential for studying adaptive immunity in clinical contexts.

\section{Results}
\subsection{Data source and modeling strategy}

\begin{figure}[!h]
\begin{center}
\includegraphics[width=.8\linewidth]{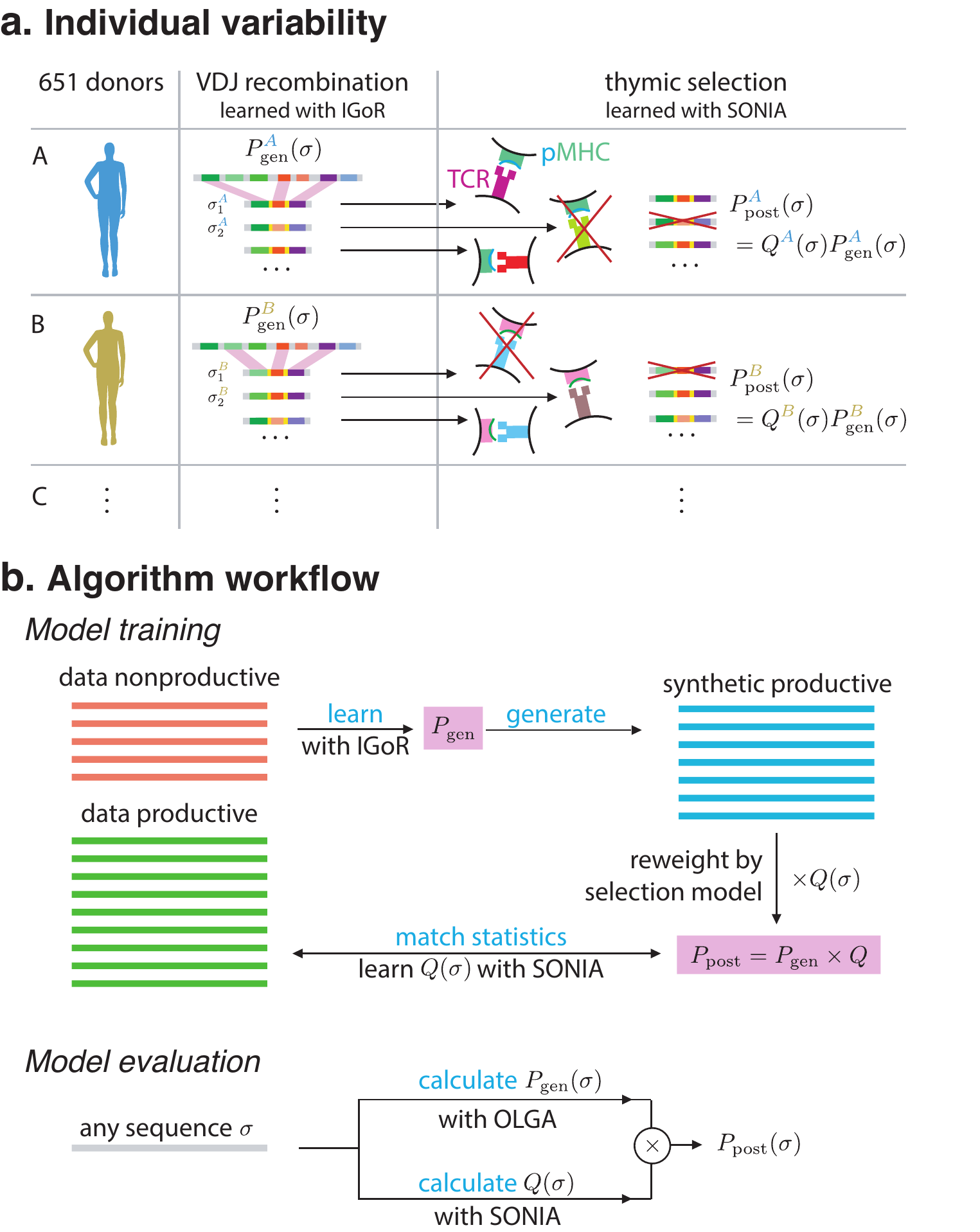}
\caption{Analysis pipeline. (A) We analyzed data from T-cell receptor beta (TRB) repertoires of 651 donors collected by Emerson et al. \cite{Emerson2017}. For each person $i = A, B, C, ...$ we define a personalized TRB generation model $P^i_{\rm gen}$, and a personalized thymic selection model $Q^i(\sigma)$, as both processes are expected to vary across individuals as a function of their genetic background, in particular their HLA type.
The generation model allows us to evaluate the probability of generating each receptor sequence $\sigma$ in each individual $i$. $Q^i(\sigma)$ tells us how likely a given receptor amino-acid sequence $\sigma$ is to pass thymic selection in a given individual. Combined together, the two models give the probability of a given TRB amino acid sequence in the repertoire of a given person $P^i_{\rm post}(\sigma)=Q^i(\sigma)P^i_{\rm gen}(\sigma)$.
(B) To learn these models, for each individual we separated sequences into productive and nonproductive sequences. Nonproductive sequences are free of selection effects and were used to learn the generation model, $P_{\rm gen}$, using the IGoR software ~\cite{Marcou2018}. Most productive sequences are subject to selection and were used to learn the selection model, $Q$, by matching the statistics of the data with those of sequences generated synthetically with $P_{\rm gen}$ (using the OLGA software \cite{Sethna2019}) and weighted by $Q$. Once the model is learned, the probabilities of amino-acid TRB sequences pre- and post-selection can be calculated using OLGA and SONIA.
}
\label{cartoon}
\end{center}
\end{figure}

We analyzed previously published RepSeq data from a large cohort study~\cite{Emerson2017} consisting of TRB nucleotide sequences from blood samples of 651 healthy individuals. Sample sizes ranged from 50,000 to 400,000 unique CDR3 amino acid beta chains. 
For each individual $i$, we learned an individual-specific generation model, which describes the probability of generating a given amino-acid sequence $\sigma$ by VDJ recombination, $P_{\rm gen}^i(\sigma)$, and an individual-specific selection model, $Q^i(\sigma)$, defined as the fitness of each sequence upon thymic selection. The resulting probability distribution of receptor sequences is $P_{\rm post}^i(\sigma)=Q^i(\sigma)P_{\rm gen}^i(\sigma)$ (Fig.~\ref{cartoon}A).
To learn these models, TRB nucleotide sequences were divided into productive and non-productive sequences, where productive sequences are defined as being in frame with no stop codon. The pipeline is summarized in Fig.~\ref{cartoon}B. We applied the IGoR algorithm~\cite{Marcou2018} to non-productive TRBs of each individual to learn $P^i_{\rm gen}(\sigma)$. Productive sequences were used to learn individual-specific selection models $Q^i(\sigma)$ as in Ref.~\cite{Elhanati2014}, by comparing them with simulated productive sequences generated from the individual specific generation models. A new software package, SONIA, was developed to perform the $Q^i(\sigma)$ inference. For each sequence in each individual the algorithm computes two probabilities:  its generation probability, $P^i_{\rm gen}$, and its post-selection probability in the periphery, $P^i_{\rm post}$. We then use them to estimate the intra- and inter-person variability.

\subsection{Individual variability of VDJ recombination statistics}

\begin{figure*}
\begin{center}
\includegraphics[width=.8\linewidth]{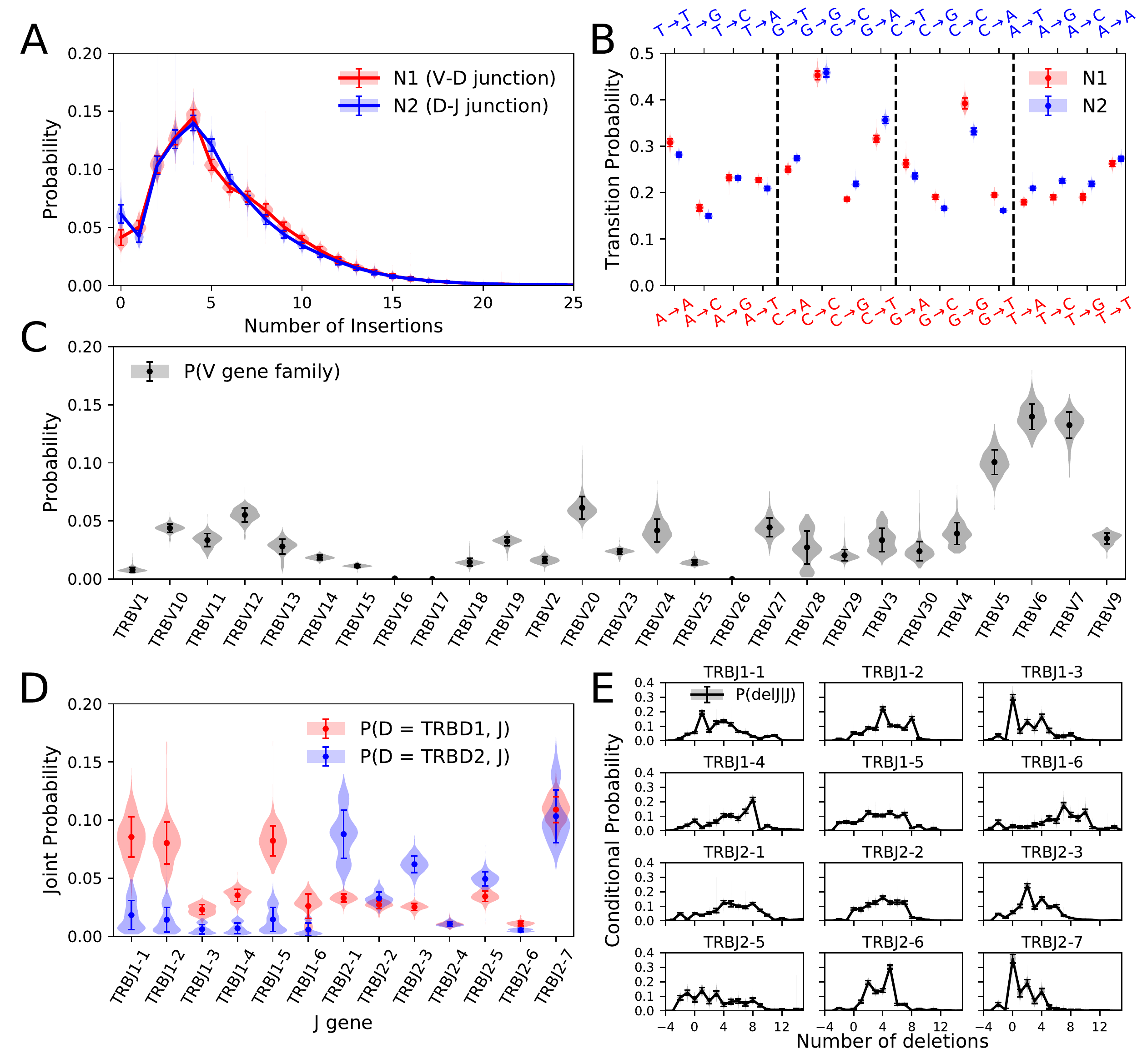}
\caption{Distribution of the individual $P^i_{\rm gen}$ model parameters over 651 individuals. All plots are violin plots with the mean and standard deviation shown by error bars. (A) Insertion length distributions of the N1 and N2 junctions. (B) Markov transition probabilities for the inserted nucleotide identities at the N1 (red) and N2 (blue) junctions. The N2 transition probabilities are organized in a reverse complementary fashion to the N1 transition probabilities.  (C) V gene family usages. (D) Joint D and J gene usages. (E) Deletion profiles for individual J genes.}
\label{generation_multipanel}
\end{center}
\end{figure*}

The model of VDJ recombination, $P_{\rm gen}$, assigns a probability to each VDJ recombination scenario \cite{Murugan2012}, where a scenario is a particular choice of the various recombination events: germline gene choice (V, D, and J), the number of deletions to those germline genes at the V-D and D-J junctions, and the number and identities of the untemplated, inserted nucleotides at each of the junctions (called N1 for the V-D junction, and N2 for the D-J junction). A detailed description of the model is given in the Methods section. 
Each recombination scenario determines a particular nucleotide sequence. The generation probability of a sequence is then the sum of all recombination scenarios that result in that sequence. Since the scenario is a hidden variable of the observed nucleotide sequence, we can use the Expectation Maximization algorithm to infer the maximum-likelihood estimator of the model parameters~\cite{Murugan2012} using IGoR~\cite{Marcou2018}.

Productive sequences then translate into an amino-acid sequences $\sigma$, and we denote by $P_{\rm gen}(\sigma)$ the probability of generation of $\sigma$ conditioned on it being productive, equal to the sum of the generation probabilities of all possible nucleotide variants divided by the probability to generate a productive sequence (an abuse of notation relative to the strict definition of $P_{\rm gen}$ with no conditioning on being productive).

We find that the generation models learned from different individuals in our cohort, $P_{\rm gen}^i$, are consistently similar to each other, with more variation in the gene usage than in the junctional diversity statistics (Fig.~\ref{generation_multipanel}). The distributions of the number of inserted N1 and N2 nucleotides vary little (Fig.~\ref{generation_multipanel}A). The biases of the untemplated inserted nucleotides, governed by a Markov model where the choice of each inserted base pair depends stochastically on the previous insertion \cite{Murugan2012}, is also conserved across individuals (Fig.~\ref{generation_multipanel}B). Note that these probabilities are also similar for the N1 and N2 insertions provided that N2 is read in the anti-sense. Likewise, gene specific deletion profiles have very low variability  (Fig.~\ref{generation_multipanel}E).
By contrast, gene usage shows greater yet moderate inter-individual variability (Fig.~\ref{generation_multipanel}C-D). Overall, these results confirm the large level of reproducibility of the generation process over a large cohort.

\begin{figure*}
\begin{center}
\includegraphics[width=.9\linewidth]{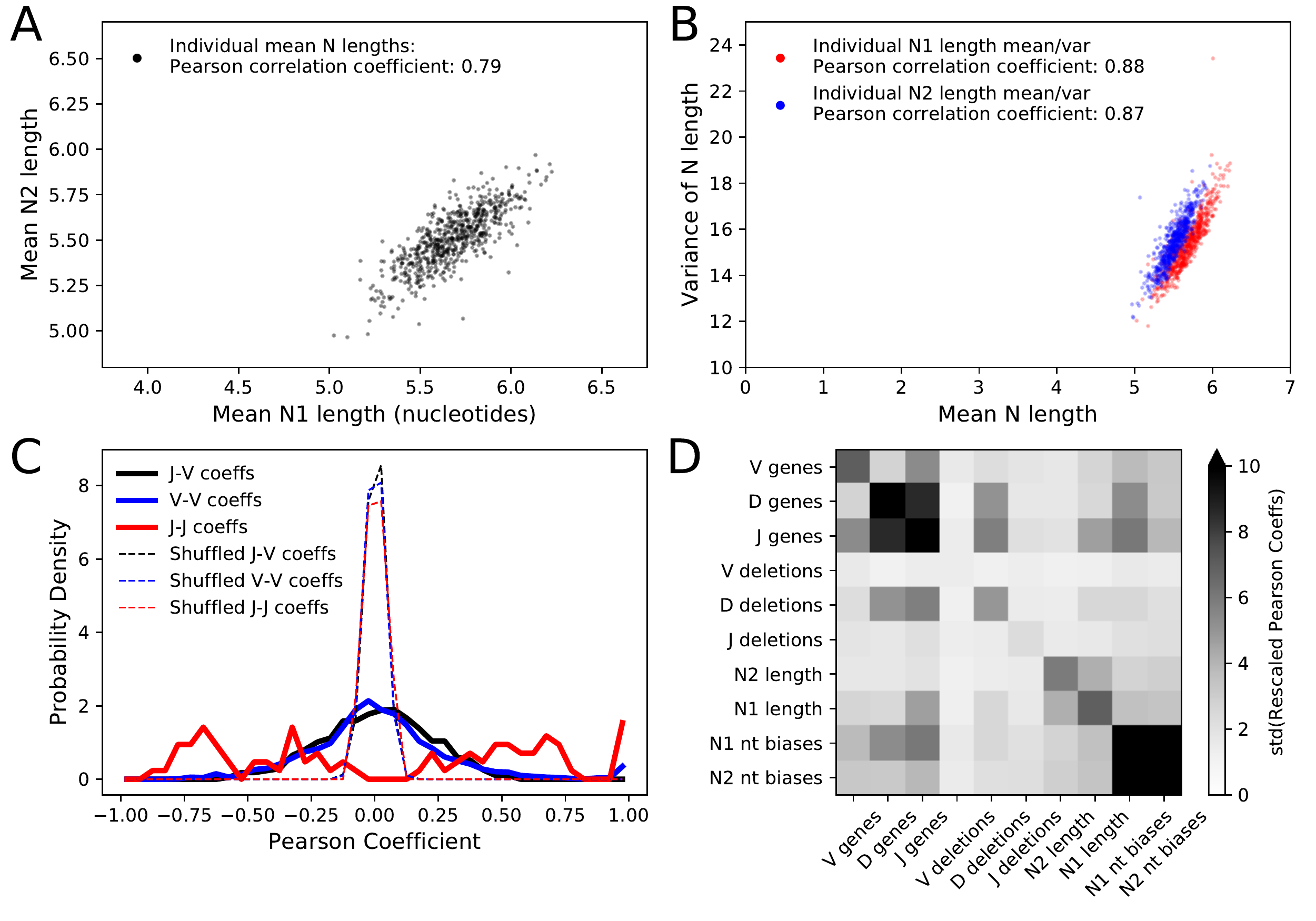}
\caption{
Correlations between model parameters across individuals. (A) Mean number of N1 (VD junction) versus N2 (DJ junction) insertions (each point is an individual). (B) Variance (across sequences) versus mean number of insertions at both junctions (each point is an individual). (C) Distribution of Pearson correlation coefficients between any two usage probabilities $P(V)$ or $P(J)$ across individuals. (D) Rescaled standard deviation of Pearson coefficients of parameter combinations over various recombination events. Values are rescaled by the standard deviation of the shuffled distribution ($\approx 0.39$ in all cases).}
\label{Covariance}
\end{center}
\end{figure*}

We then asked whether these small individual variations in the recombination statistics were correlated as a result of shared biological mechanisms or genetic factors.
We found that the numbers of insertions at the two junctions were highly correlated with each other (Pearson's $r=0.79$), meaning that individuals that tend to have longer N1 insertions also tend to have longer N2 insertions on average (Fig.~\ref{Covariance}A).  
N1 insertions were also slightly longer by $\sim0.17$ insertions on average.
The variance of the number of insertions calculated over the repertoire of one individual is extremely correlated to its mean (Pearson's coefficients of 0.88 and 0.87, Fig.~\ref{Covariance}B), suggesting a single individual-specific parameter controlling both N1 and N2 length distributions. This parameter is likely linked to the activity of the Terminal Deoxynucleotidyl Transferase (TdT) enzyme responsible for N insertions \cite{Bogue1992}.

To quantify other correlations we calculated Pearson's correlation coefficient over the population between combinations of various parameters. In order to determine significance and account for the finite cohort size we also compute a \lq shuffled\rq \ Pearson's coefficient for each parameter combination by scrambling the individuals to destroy correlations. Fig.~\ref{Covariance}C shows the normalized distribution of Pearson's correlation for the combinations of the marginal distributions of V, D, and J usages. Correlations between $V-V$, $J-J$, and $V-J$ marginals all show substantial excess of positive and negative values relative to the shuffled control.
Full parameter co-variations are shown in Figs.~S1-S3.
To determine which types of parameters co-vary the most, we computed the rescaled standard deviation of the Pearson's correlation coefficients of all combinations of parameter types (Fig.~\ref{Covariance}D). This analysis reveals that V gene usage co-varies with itself, D and J usages are also correlated with each other, as well as N1 length with N2 length, and the insertion biases at N1 and N2 with each other.

\subsection{Learning models of thymic selection with SONIA}
\begin{figure*}
\begin{center}
\includegraphics[width=0.9\linewidth]{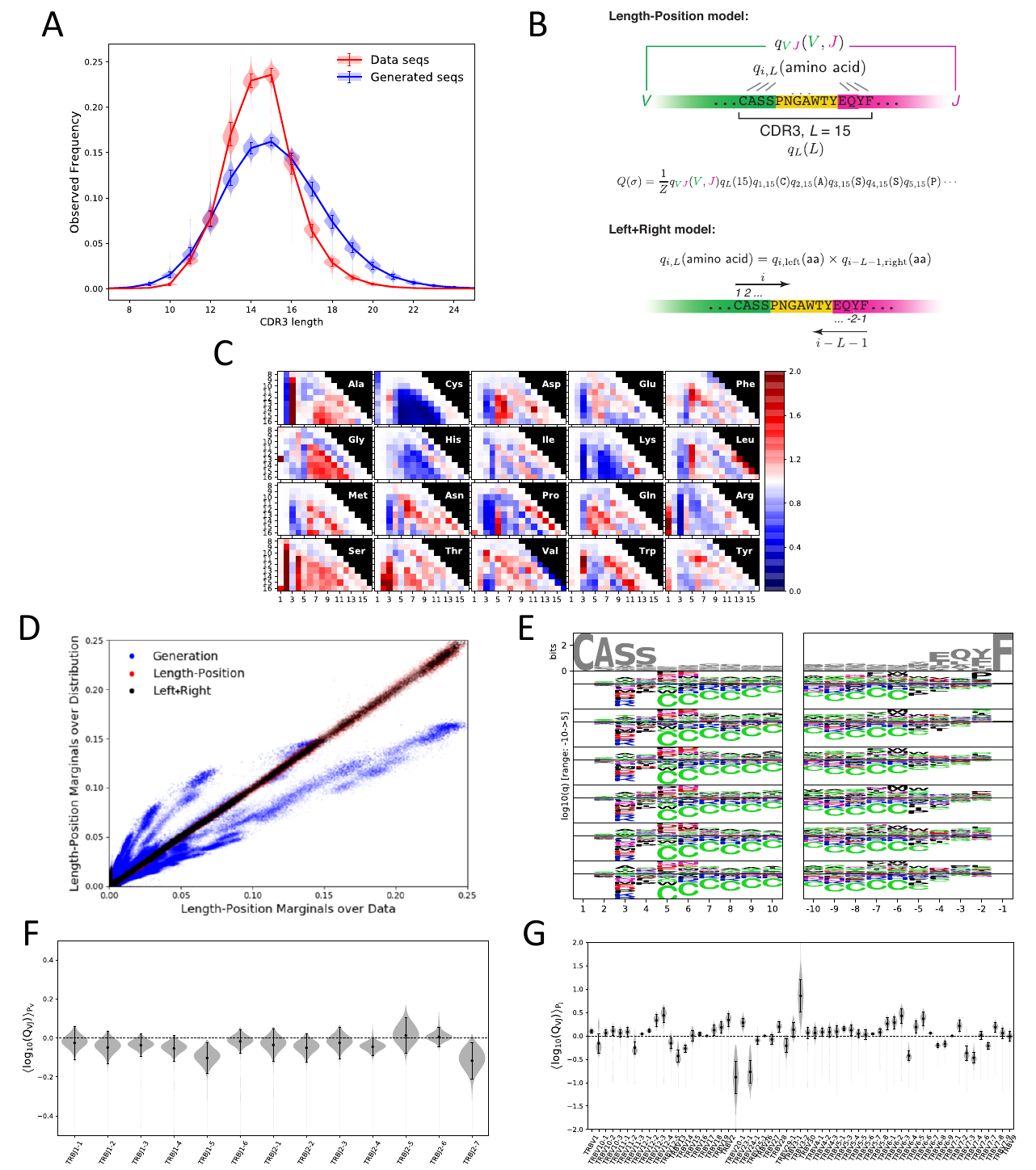}
\caption{Thymic selection models of 651 individuals.
  (A) Length distribution of the Complementarity Determining Region 3 (CDR3) of TRB before (as predicted by the $P_{\rm gen}$ model, in blue), and after (data, in red) thymic selection. Violin plots show variability across individuals.
  (B) Schematic of the two SONIA model architectures used in this article. Both models have selection factors for the joint choice of $V$ and $J$, $q_{VJ}$, and for the CDR3 length $L$, $q_L$. The \textit{LengthPosition} model has selection factors defined for each amino acid at each position $i$ and length $L$, $q_{i,L}$. The \textit{Left+Right} model factorizes those factors into two contributions depending on the position of the amino acid from the left and right, respectively.
  (C) Amino-acid selection factors $q_{i,L}$ of the \textit{LengthPosition} model as a function of position $i$ and $L$ for each of the 20 amino acids. These factors are consistent with previous reports on a smaller cohort \cite{Elhanati2014}.
  (D) Model prediction for the frequencies of all features of the \textit{LengthPosition} model  (V,J joint usage, CDR3 length, and amino acid usage at each position and length).
  The \textit{Left+Right} model reproduces all the probabilities despite not having learned them directly.
  (E) Model parameters of the \textit{Left+Right} model, for right ($\log_{10} q_{i,\rm left}({\rm aa})$) and left ($\log_{10} q_{i,\rm right}({\rm aa})$) displayed as sequence logos for 6 individuals. The first row shows the sequence logos for the amino acid usage from the generation model alone (consistent across individuals), with the usual convention that the total height of the logo is equal to the Shannon entropy of amino acid usage at this position, and the relative height of each letter is proportional to its usage.  (F,G) Distributions of selection factors for V and J genes, $q_{VJ}$, over the population (averaged over one of them, as selection factors as defined for the joint usage of V and J).}
\label{selection}
\end{center}
\end{figure*}

After VDJ recombination, new T cells go through an initial selection process in the thymus before being released as naive T cells to the periphery. Positive thymic selection selects for functionally useful receptors, while negative selection removes T cells that recognize self-peptides to avoid auto-immunity. Thymic selection skews the statistics of the repertoire of TRB sequences in quantifiable ways. This can be seen by comparing the length distribution of the Complementarity Determining Region 3 (CDR3, running from a conserved cysteine near the end of V segment through a conserved phenylalanine near the beginning of the J segment) of productive sequences drawn from the generation model to observed sequences (Fig ~\ref{selection}A). We observe a substantial narrowing of the distribution post-selection, eliminating sequences much longer or shorter than 14-15 amino acids~\cite{Elhanati2014}.

To characterize these differences more systematically, we use a statistical model of selection to account for differences between the  repertoire generated from the raw VDJ recombination (pre-selection) and the observed repertoire of productive sequences (post-selection). Since selection acts on the functionality of a receptor we restrict ourselves to productive amino acid sequence statistics.
Mathematically, we require that the post-selection distribution, $P_{\rm post}=Q(\sigma)P_{\rm gen}(\sigma)$, agrees with the statistics of productive sequences in the frequency of a select set of features, $f \in\mathcal{F}$, while remaining as close as possible to $P_{\rm gen}$ (where distance is measured by the Kullback-Leibler divergence, $\sum_\sigma P_{\rm post}(\sigma)\ln (P_{\rm post}(\sigma)/P_{\rm gen}(\sigma))$). {This is done by choosing the sequence-specific selection factors $Q(\sigma)$ which can be shown to take the form (see Methods):}
\beq\label{eq:Q}
Q(\sigma) =\frac{1}{Z}\prod_{f\in \mathcal{F}(\sigma)} q_f,
\eeq
where $\mathcal{F}(\sigma)\subset \mathcal{F}$ is the subset of features present in sequence $\sigma$.
Solving for the factors $q_f$ that match the frequencies of features in the data is equivalent to maximum likelihood estimation (MLE).

Features may be the presence of a given amino-acid at a given position, the use of a particular V or J gene, a particular CDR3 length, or any combination thereof. For example, some of the features of the TRB designated by (CASSGRQGVATQYF, TRBV06-05, TRBJ02-05) are `CDR3 length 14', `S in position 2 from the left',  `Y in position -2 from the right' and `V gene is TRBV06-05'.

To facilitate the definition and learning of such selection models, we introduce the software package SONIA. SONIA allows for a flexible definition of model features and infers the selection factors $q_f$ using MLE. The input to SONIA is a list of selected amino acid sequences and, if needed, their V and J gene choice. By default SONIA uses $P_{\rm gen}$ as provided by an IGoR inferred model (using OLGA as a generation engine~\cite{Sethna2019}), but it can also take as an input a custom sample of pre-selection sequences. This can be useful for identifying selection pressures during immune challenges using different choices of pre and post-selection repertoires (see Methods for details).

We applied SONIA using two models corresponding to two choices of feature sets. In the \textit{LengthPosition} model \cite{Elhanati2014}, features include all possible choices of combinations of V and J genes, all possible CDR3 lengths, as well as amino acids usage at each position and length (Fig.~\ref{selection}B, top). This choice allows for great flexibility at the cost of many parameters. The  \textit{LengthPosition} model replicates the results of Ref.~\cite{Elhanati2014} (Fig.~\ref{selection}C).

The number of parameters can be reduced by noting that selection pressures on amino acids near the 5' (left) or 3' (right) end of the CDR3 appear to depend only on their relative position to that end, regardless of CDR3 length (Fig.~\ref{selection}C).
The \textit{Left+Right} model exploits that regularity by defining features of amino-acid usage at positions relative to the 5' end of the CDR3 (denoted by a positive index), or to its 3' end (denoted by a negative index). This model has much fewer parameters, since features are defined for left and right positions regardless of CDR3 length, and can be written as a special parametrization of the {\em LengthPosition} model, in which each amino acid contributes to the selection factor through the product of a left and a right factor (Fig.~\ref{selection}B, bottom).

To evaluate the accuracy of the \textit{Left+Right} model, we computed its predictions for the frequencies of amino acid usages at each position and length (Fig.~\ref{selection}D, see also Fig.~S4 for overall amino-acid usage). These statistics are by construction matched by the \textit{LengthPosition} model but not necessarily by the \textit{Left+Right} model, and thus provide a good test of the validity of the parameter reduction it affords. 
While predictions from VDJ generation model (blue dots) do not reproduce the empirical frequencies well, highlighting the need of a selection model, both the \textit{LengthPosition} (red dots) and the \textit{Left+Right} (black dots) models match the data well.
As the \textit{Left+Right} model captures the observed behavior with fewer parameters, we will work with this model for the remainder of the paper.

Fig.~\ref{selection}E displays the selective pressures on the CDR3 amino acid composition ($q_{i,{\rm left}}$ and $q_{i,{\rm right}}$) from the left and right positions across a choice of 6 (out of 651) individuals, in the form of sequence logos. These selective factors are mostly conserved across individuals.
Fig.~\ref{selection}F and Fig.~\ref{selection}G show the selection factors for the V and J genes ($q_{VJ}$) averaged over one of the two segments. Again, the pattern is mostly concordant across the population, but with some substantial differences for a few genes that have greater variability. Thus, much as in the generation process, individual variability in the selection process is moderate and concentrates on gene usage rather than CDR3 statistics.

\subsection{Population variability}

\begin{figure*}
\begin{center}
\includegraphics[width=\linewidth]{Fig5.pdf}
\caption{
(A-C) Distributions of $P_{\rm gen}$, $Q$, and $P_{\rm post}$ calculated over many sequences for each individual. Shown are the post-selection productive TRBs from each individual (red), and pre-selection sequences generated from the individual's VDJ generation model $P_{\rm gen}^i$ (blue). The distributions for all individuals are visualized using a density map indicating the local density of probability distribution curves over the cohort. (D) Density maps of the model distributions for the VDJ recombination scenarios, $P_{\rm scenario}$, the nucleotide sequences, $P_{\rm gen}^{\rm nt}$, the productive amino-acid sequences upon generation, $P_{\rm gen}$, and post-selection amino-acid sequences $P_{\rm post}$, over the population. The same convention for the density map is used. Error bars for (A-D) are the standard deviation over the population. (E) Distributions of the Shannon entropies of $P_{\rm scenario}$, $P_{\rm gen}^{\rm nt}$, $P_{\rm gen}$, and $P_{\rm post}$ over the population.
  (F) Mean vs variance of $\log_{10} P_{\rm gen}$  and $\log_{10} P_{\rm post}$ over both productive and generated sequences for each individual. The linear relation suggests a single parameter explaining variability in the population.}
\label{pgen_q_ppost_dists}
\end{center}
\end{figure*}
To quantify more precisely the variability of the generation and selection processes across 651 individuals, we computed the distributions of $\log_{10}P_{\rm gen}$, $\log_{10}Q$, and $\log_{10}P_{\rm post}$ for each individual (see Methods). Figs.~\ref{pgen_q_ppost_dists}A-C show the results as a density map over the entire population, indicating strong consistency between individuals. The distributions over sequences from the model (obtained by sampling from $P_{\rm post}$ using importance sampling, black curve) agree very well with those obtained from the data (red). By contrast, sequences generated from $P_{\rm gen}$, without selection factors (blue) fail to reproduce the data.

The shift to high $Q$ values from the pre- to the post-selection model is present by construction in the distribution of the $Q$ (Fig.~\ref{pgen_q_ppost_dists}B), because the post-selection ensemble should be enriched in high selection factors. However, a similar shift to higher probabilities from pre- to post-selection is indicative of a correlation between the generation probability, $P_{\rm gen}$, and the selection factor, $Q$ (Table \ref{tab:intraindiv}, Fig.~S5). This correlation suggests that evolution has shaped VDJ recombination to favor sequences that are likely to pass thymic selection, as previously argued \cite{Elhanati2014}.

Fig.~\ref{pgen_q_ppost_dists}D summarizes the distributions of probabilities $P$ in different probability ensembles of decreasing diversity: raw VDJ recombination scenarios (black), generated nucleotide sequences (green), pre-selection productive amino acid sequences (blue, same as the blue curves in Fig.~\ref{pgen_q_ppost_dists}A), and post-selection productive amino acid sequences (red, the mean of which is the black curve in Fig.~\ref{pgen_q_ppost_dists}C). The negative of the mean of $\log_{10}(P)$ is, up to a $\ln(2)/\ln(10)$ factor, equal to the Shannon entropy of the distribution expressed in bits, $\<-\log_2(P)\>_P$. Fig.~\ref{pgen_q_ppost_dists}E shows the distribution of these entropies across the population. The width of the distributions of $\log_{10} P$ is strongly correlated with their means across individuals and also from pre-selection to  post-selection (Fig.~\ref{pgen_q_ppost_dists}F), suggesting again a single parameter driving individual variability, possibly the average number of N insertions.

\begin{figure*}
\begin{center}
\includegraphics[width=.9\linewidth]{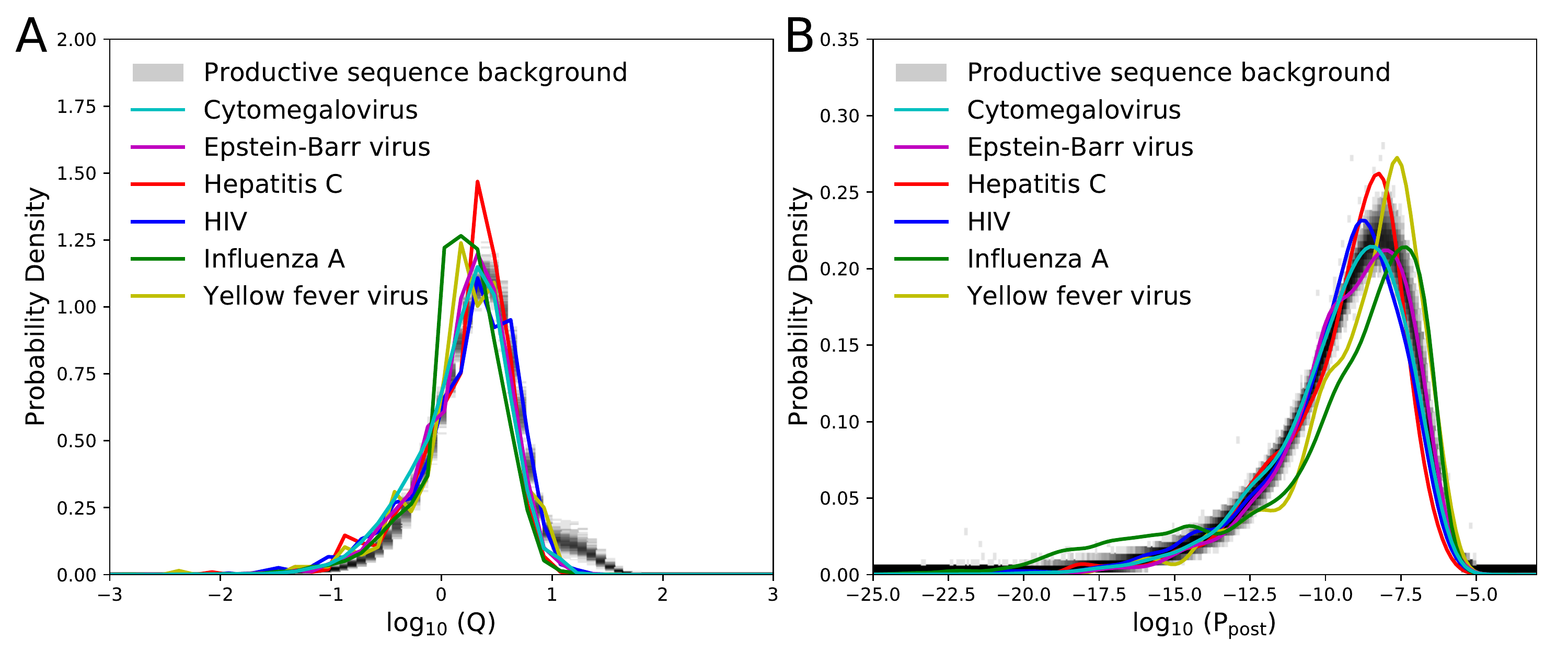}
\caption{Distribution of TRB sequences from the VDJdb database specific to human viruses \cite{Bagaev2019} compared to the productive sequences from the blood of 651 individuals.  A) $\log_{10}  (Q^{\rm univ})$ distribution for each individual's productive data sequences (gray heatmap) and for viral-specific TCRs from the VDJdb database. B) $\log_{10}(P_{\rm post}^{\rm univ})$ distributions. The VDJdb $\log_{10}(P_{\rm post}^{\rm univ})$ distributions are Gaussian-smoothed for clarity. $Q^{\rm univ}$ and $P_{\rm post}^{\rm univ}$ are `universal' models learned from sequences randomly drawn from all individuals.}
\label{VDJdb_fig}
\end{center}
\end{figure*}

We also plot the $P_{\rm post}$ distributions of TRBs from the VDJdb database that are known to be specific to human viruses \cite{Bagaev2019} (Fig.~\ref{VDJdb_fig}). There does not appear to be a substantial shift in the post-selection probability of these viral-specific sequences as compared to productive TRBs from blood. A similar absence of bias was previously reported for the distribution of generation probabilities \cite{Sethna2019}, suggesting that the VDJ recombination process is not explicitly skewed towards generating these viral-specific sequences. Our results further show that thymic selection also does not seem to be biased to select for sequences specific to these viral epitopes.

\subsection{Quantifying overall variability and its contribution due to generation and selection}

\begin{table}
\begin{center}
\caption{Intra-individual variation}
\label{tab:intraindiv}
\begin{tabular}{|c|c|c|c|}
\hline
Seqs & Quantity & Intra-indiv Var & \% \\ 
\hline
Gen seqs &Var  $\log_{10} P_{\rm gen}$ & 7.05 $\pm$ 0.49 & 78.2\% \\
&Var $\log_{10} Q$ & 0.407 $\pm$ 0.016 & 4.52\% \\
&Cov $(\log_{10} P_{\rm gen}, \log_{10} Q)$ &0.778 $\pm$ 0.069 & $2\times$8.64\%\\
&Var  $\log_{10} P_{\rm post}$ &9.01 $\pm$ 0.63 & 100\%\\
\hline
Data seqs &Var  $\log_{10} P_{\rm gen}$ & 4.67 $\pm$ 0.53 & 86.9\% \\
&Var $\log_{10} Q$ & 0.185 $\pm$ 0.009 & 3.44\% \\
&Cov $(\log_{10} P_{\rm gen}, \log_{10} Q)$ &0.258 $\pm$ 0.033 & $2\times$4.81\%\\
&Var  $\log_{10} P_{\rm post}$ &5.37 $\pm$ 0.56 & 100\%\\
\hline
\end{tabular}
\end{center}
\end{table}

The overall variability in the TRB repertoire can be characterized both between and within individuals in the population, by calculating the variance of the distribution of $\log_{10}P_{\rm post}$, which gives a measure of the typical fold-variation. Since $\log_{10}P_{\rm post}(\sigma)=\log_{10}P_{\rm gen}(\sigma)+\log_{10}Q(\sigma)$, this variance can be decomposed as:
\begin{equation}
  \begin{split}
  \mathrm{Var}(\log_{10} P_{\rm post})=&\mathrm{Var}(\log_{10} P_{\rm gen})+\mathrm{Var}(\log_{10} Q)\\
  &  +2\mathrm{Cov}(\log_{10} P_{\rm gen},\log_{10} Q).
  \end{split}
  \end{equation}

To quantify the range of repertoire variability within an individual we calculate the variances and covariance of $\log_{10}P_{\rm gen}^i$, $\log_{10}Q^i$, $\log_{10}P_{\rm post}^i$ over the data sequences, and synthetic sequences for each individual. Table~\ref{tab:intraindiv} summarizes the average of these variances over the 651 individuals. 80\% of the variation comes from the generation process, with the remainder mostly stemming from a strong correlation between selection and generation, as previously discussed (Fig.~\ref{pgen_q_ppost_dists}A and C, SI Fig.~5).

Variations in the probabilities of given sequences across individuals (averaged over sequences, see Methods for details) are much lower (Table~\ref{tab:interindiv}), highlighting the high level of consistency in the population. The total variance of 0.091 in $\log_{10} P_{\rm post}$ corresponds to relative variations of $10^{\pm \sqrt{0.091}}\in (-50\%,+100\%)$ in the probability of sequences. While those differences are substantial in absolute terms, they are $1.6\%$ of the variance over sequences within an individual ($\approx 5.4$, see Table~\ref{tab:intraindiv}). Much of this variance again stems from VDJ generation.

\begin{table}
\begin{center}
\caption{Inter-individual variation}
\label{tab:interindiv}
\begin{tabular}{|c|c|c|c|}
\hline
Seqs & Quantity & Inter-indiv Var & \% \\ 
\hline
Gen seqs & Var  $\log_{10} P_{\rm gen}$ & 0.121 & 95.5\% \\
& Var $\log_{10} Q$ & 0.0175 & 13.7\% \\
& Cov $(\log_{10} P_{\rm gen}, \log_{10} Q)$ &-0.00586 &$2\times$ -4.62\% \\
& Var  $\log_{10} P_{\rm post}$ &0.127 & 100\% \\
\hline
Data seqs & Var  $\log_{10} P_{\rm gen}$ & 0.0792 &87.0\% \\
& Var $\log_{10} Q$ & 0.0132 & 14.5\% \\
& Cov $(\log_{10} P_{\rm gen}, \log_{10} Q)$ &-0.000697 &$2\times$-0.766\% \\
& Var  $\log_{10} P_{\rm post}$ &0.0910 & 100\% \\
\hline
\end{tabular}
\end{center}
\end{table}

\begin{figure}
\begin{center}
\includegraphics[width=\linewidth]{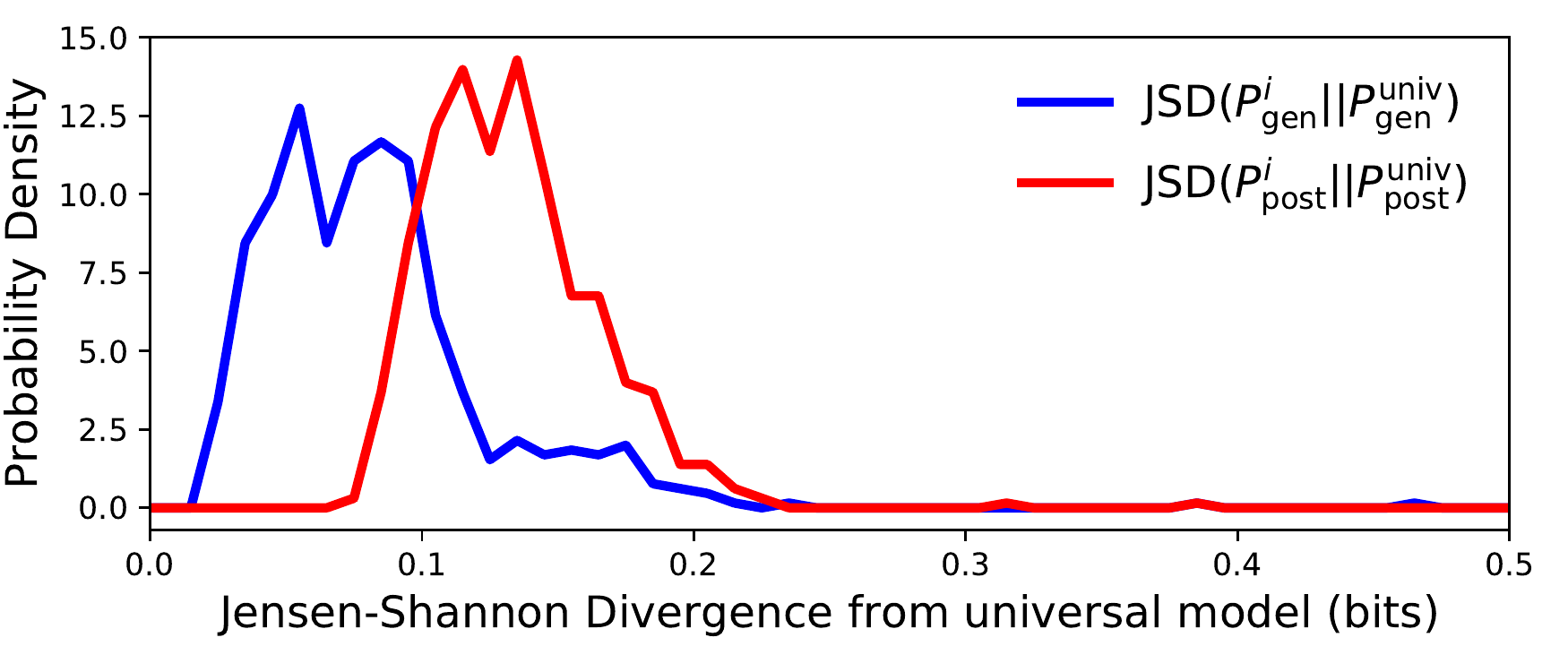}
\caption{Normalized distributions of the Jensen-Shannon divergence (JSD) of each individual from the universal model for both $P_{\rm gen}$ and $P_{\rm post}$.}
\label{info_theory_divs}
\end{center}
\end{figure}

To further characterize variability, we learned `consensus' or `universal' models from sequences sampled randomly from each individual. To this end we inferred a consensus VDJ generation model ($P_{\rm gen}^{\rm univ}$) from out-of-frame sequences, and a consensus \textit{Left+Right} SONIA model  ($Q^{\rm univ}$ and $P_{\rm post}^{\rm univ}$) from the productive sequences (Methods). We then compared each individual model to the universal model using the Jensen-Shannon divergence, an information-theoretic measure of distance between probability distributions expressed in bits and directly comparable to entropies (Methods). The distributions of JSD($P_{\rm gen}^i,P_{\rm gen}^{\rm univ}$) and JSD($P_{\rm post}^i,P_{\rm post}^{\rm univ}$) over the cohort highlight the consistency of these models with most individuals having $<0.3$ bits JSD from both $P_{\rm gen}^{\rm univ}$ and $P_{\rm post}^{\rm univ}$ (Fig.~\ref{info_theory_divs}). This should be compared to the associated entropies of $>30$ bits for either distribution (Fig.~\ref{pgen_q_ppost_dists}E).

\section{Discussion}
By applying distinct computational procedures to the nonproductive and productive sequences of the TCR repertoires of a large cohort of 651 donors, we were able to learn individual-specific models of repertoires, separating the processes of generation and thymic selection. This allowed us to quantify precisely the variability of each process within the population.

We found that the TRB generation process varied only moderately between individuals, with two main drivers: gene usage and average length of untemplated insertions. Because insertions contribute a lot to the generation probability, the latter is the main driver of variability in the distribution of $P_{\rm gen}$ itself. V,D, and J gene usage variability may be due to variations in the regulatory signals, both genetic and epigenetic, that control the operation of the Recombination-Activating Gene (RAG) protein that initiates the recombination process \cite{Schatz2011}. It may also be due to variations in gene copy numbers of the gene segment, as was observed in the related case of the IgH locus \cite{Luo2016a}. Variations in the mean number of insertions could be attributed to differences in expression of TdT as well as other proteins involved in the non-homologous end joining pathway \cite{Lieber:2010in}.

We found that the inferred selection models were also variable between individuals, but the magnitude of these variations remains limited, which may be surprising considering that different individual's repertoires are subject to different selective pressures due to diverse HLA backgrounds. Overall, the ratio of the total variability across individuals to the variability across sequences within an individual (measured by the variance of the logarithm of the sequence probability) was only 1.6\%, of which about 85\% came from variations in the generation process, and 15\% from the selection process. We also found that sequences that were previously identified to be specific to human viruses did not differ in their generation or selection probability from generic sequences from blood, finding no evidence in our models for an evolutionary mechanism to favor such viral-specific sequences (as suggested in \cite{Thomas2019}), neither in the process of VDJ recombination, nor through thymic selection.

Thymic selection of naive T cells was found to be well captured by a model where selection acts independently on each amino acid, regardless of the sequence context. The variability in the inferred parameters for the selection models is not large in the population, identifying reproducible features in different individuals. This suggests that the main statistical effects of thymic selection captured by our model are mostly universal, probably driven by positive selection for amino acids that makes a folding functional receptor. The effect of HLA specific positive and negative selection, on the other hand, might not be well captured by this kind of a model, which focuses on finding broad sequence features rather than specific sequences to harness more statistical power, although variations in the V and J selection factors may reflect HLA types.
Our approach thus complements the strategy of looking for associations of particular TCRs with HLA type, which was previously applied to the same dataset \cite{DeWitt2018}. An obvious limitation of this and other studies of that dataset is that it comprises a restricted subset of the human population. 

While in this study we used SONIA for the purpose of comparing peripheral to pre-selection repertoires, the software is written to be flexible in several ways. First, if can be used to infer selection factors between any two repertoires (observed or generated), by inferring selection factors that match the statistics of the two samples. Second, SONIA can go beyond selection pressures on single amino acids, allowing features of pairs or motifs of amino acids. Finally, SONIA can be applied to other chains than TRB, notably the alpha chain of the TCR (TRA) as well as immunoglobulin IgH.

SONIA's flexibility opens up the possibility of using SONIA to find statistical correlations in various biological or clinical contexts. SONIA could be applied to samples that are known to have responded to some perturbations, for example after vaccination or infection \cite{Thomas2014b,Pogorelyy2018}. In such a context clone sizes may be crucial to identify the underlying changes. To facilitate this, SONIA can also infer selection factors from read-count weighted repertoires.
A major challenge in the field of immune repertoire profiling remains to decipher the specificity of the TCR-pMHC interaction. Vaccine design, immunotherapy and therapy for autoimmune conditions would all greatly benefit from the ability to find or design TCRs with known specificity. In the last couple of decades experimental methods have been developed for identifying TCRs specific to given antigens~\cite{wolfl2007, Altman94,Glanville2017, Dash2017}. Based on accumulated TCR binding data \cite{Bagaev2019}, computational methods have been proposed recently that can find clusters of similarly reactive TCRs~\cite{Glanville2017, Dash2017,Pasetto2016,Pogorelyy2018}, or to predict TCR specificity to a given epitope using machine-learning techniques \cite{Jurtz2018,Sidhom2018,Springer2019,Jokinen2019}. SONIA could be used to learn flexible models of these antigen-specific TCR subsets and to study their organization. It could also be applied to identify specific selective pressures in particular subsets, defined by HLA specificity, pathogenic history, clinical status, T-cell phenotype (naive, effector, memory, CD4, CD8, regulatory T cells), or to differentiate distinct samples from the same individual, such as blood, tissue, or tumor samples.

\section{Methods}

\subsection{Data}
The data used for the inference of both the VDJ generation models and the subsequent selection models are the Adaptive Biotechnologies sequenced TRB repertoires of Emerson et. al.~\cite{Emerson2017}. An initial quality control pass was done over the 664 individuals to ensure at least 10,000 unique out of frame sequences to be used to infer the VDJ generation model. 651 individuals passed this threshold and all were used in the subsequent analyses.

All analyses were done on unique nucleotide reads, discarding any cell count information. This is done to ensure that each sequence is reflective of a single recombination event, which is an important restriction when modeling VDJ recombination and thymic selection. For some selection modeling purposes (e.g. modeling antigen exposure), cell counts may be incorporated.

In practice, amino-acid sequences are reduced to the choice of V and J, and the full amino acid CDR3 sequence.

Sequences were determined to be productive and used in the selection analysis if they had a non-zero $P_{\rm gen}$. Beyond being an in-frame sequence without stop codons, this requires that a sequences retains the conserved residues defining the CDR3 region (Cysteine on the $5^\prime$ end, Phenylalanine or Valine on the $3^\prime$ end) as well as aligning to non-pseudo V and J genes.

\subsection{Generation model}
The generation model is defined at the level of the recombination scenarios in order to reflect the underlying biology of VDJ recombination. Each recombination scenario is defined by the gene choice ($V$, $D$, and $J$); deletions/palindromic insertions for each gene ($d_V$, $d_D$, $d'_D$, and $d_J$); and the sequence of non-templated nucleotides at each junction ($m_1, \ldots, m_{\ell_{VD}}$ and $n_1, \ldots, n_{\ell_{DJ}}$). The probability of a recombination scenario is given in the factorized form:
\beq\label{eq_VDJmodel}
\begin{split}
P_{\rm scenario} &= P_{\rm V}(V)  P_{\rm delV}(d_V | V)  P_{\rm DJ}(D, J)\\
&\times P_{\rm delD}(d_D, d'_D|D) P_{\rm delJ}(d_J|J)  \\
&\times P_{\rm insVD}(\ell_{\rm VD})p_0(m_1)\left[\prod_{i=2}^{\ell_{VD}} S_{\rm VD}(m_{i}|m_{i-1})\right] \\
&\times P_{\rm insDJ}(\ell_{\rm DJ})q_0(n_{\ell_{DJ}})\left[\prod_{i=1}^{\ell_{DJ}-1}S_{\rm DJ}(n_i|n_{i+1})\right].
\end{split}
\eeq
This model factorization, originally from Murugan \textit{et al}, has been shown to capture the relevant correlations between the different recombination events in TRB \cite{Murugan2012}.

The probability of a nucleotide sequence $x$ is given by:
\beq
P_{\rm gen}^{\rm nt}(x)=\sum_{{\rm scenario}\to x} P_{\rm scenario},
\eeq
and the probability of a productive amino-acid sequence is:
\beq
P_{\rm gen}(\sigma)=\frac{1}{F}\sum_{x\to \sigma}P_{\rm gen}^{\rm nt}(x),
\eeq
where $F = \sum_{{\rm scenario}|\rm prod}P_{\rm scenario}$ is the total probability that a random recombination event is productive (in-frame, no stop codons, preserves conserved residues, and does not use pseudo-genes as germline gene choices). $F$ can be computed directly from a generative model using OLGA \cite{Sethna2019}.

\subsection{Selection model}
To minimize the Kullback-Leibler distance between $P_{\rm post}$ and $P_{\rm gen}$ while enforcing the constraints $\sum_{\sigma:f\in \mathcal{F}(\sigma)}P_{\rm post}(\sigma)\equiv P_{\rm post}(f)=P_{\rm data}(f)$ for each $f$, we extremize the following Lagrangian:
\begin{equation}
  \sum_\sigma P_{\rm post}(\sigma)\left[\ln \left(\frac{P_{\rm post}(\sigma)}{P_{\rm gen}(\sigma)}\right)- \sum_{f\in\mathcal{F}(\sigma)} \lambda_f -\mu \right],
\end{equation}
where $\lambda_f$ are Lagrange multipliers constraining the frequencies of $f$, while $\mu$ ensures the normalization of $P_{\rm post}$. This extremization yields the form of $P_{\rm post}$:
\begin{equation}
  P_{\rm post}(\sigma)=P_{\rm gen}(\sigma)\exp\left(\sum_{f\in\mathcal{F}(\sigma)}\lambda_f\right).
\end{equation}
Defining $q_f=e^{\lambda_f}$, and $Z=e^{-\mu}$, we obtain Eq.~\ref{eq:Q}. Given that form, the Lagrange multipliers must be adjusted to satisfy the constraints. Doing so is equivalent to maximizing the likelihood of the data under the model:
\begin{equation}
  \mathcal{L}=\frac{1}{N}\sum_{\sigma\in {\rm data}} \ln P_{\rm post}(\sigma|\{\lambda_f\}),
\end{equation}
where $N$ is the number of data sequences. This can be shown by noting that the gradient of the log-likelihood,
\begin{equation}
  \begin{split}
    \frac{\partial\mathcal{L}}{\partial \lambda_f}&=\frac{1}{N}\left(\sum_{\sigma\in {\rm data}:f\in\mathcal{F}(\sigma)}1\right) - P_{\rm post}(f)\\
    &=P_{\rm data}(f)-P_{\rm post}(f),
    \end{split}
\end{equation}
cancels when the constraints are satisfied.
  
\subsection{SONIA implementation}
SONIA is a python software built to define and infer feature-defined selection models. SONIA has built in procedures for defining and identifying sequence features of CDR3 sequences. SONIA also ships with the prepackaged selection models of \textit{LengthPosition} and \textit{Left+Right} features. With a feature model defined, SONIA takes as an input a list of productive amino acid CDR3s, along with any aligned V/J genes. This list of observed CDR3s can be either reduced to unique sequences (useful when learning thymic selection and the background statistics are based on unique sequences) or sequences taken with their clonality to account for a non-flat clone size distribution. As an optional input, SONIA can read in baseline CDR3 and aligned V/J genes to use as the background that the selection model is learned from. Alternatively, OLGA's sequence generation machinery~\cite{Sethna2019} is built into SONIA so a generation model can be specified and background sequences automatically generated.

SONIA has built-in methods to compute the feature marginals over the data sequences, background sequences, and the selection model. These marginals are use to fit the selection model iteratively using TensorFlow keras~\cite{chollet2015keras,tensorflow2015} with the Kullback-Leibler divergence as a loss function. We checked the convergence of the algorithm and its satisfying of the constraints after convergence (Fig.~S6)

An inferred SONIA model can be used to compute overall selection factors $Q$ of any sequence. In combination with OLGA, SONIA can compute $P_{\rm post}$ and to generate selected sequences through rejection sampling. 

\subsection{Distributions of probabilities}
We produced the distributions of $P_{\rm gen}$, $Q$, and $P_{\rm post}$ shown in Fig.~\ref{pgen_q_ppost_dists}A-C by comparing the productive data sequences of each individual to a synthetic sample of productive sequences generated from $P_{\rm gen}^i$ of that individual using OLGA \cite{Sethna2019}. The number of generated sequences for each individual were matched to the number of productive data sequences. For each dataset, we calculated $P_{\rm gen}^i$ using OLGA, and $Q^i$ and $P_{\rm post}^i$ using SONIA's \textit{Left+Right} model. The $Q$-weighted curves are determined by weighting each generated sequence by its selection factor $Q^i$ and then renormalizing.

For Fig.~\ref{pgen_q_ppost_dists}D, we used 300,000 scenarios, nucleotide, and amino acid sequences were generated from each individual's VDJ generation model. Again, we used OLGA to compute the various generation probabilities $P^i$, where $P^i$ is $P^i_{\rm scenario}$, $P^{i,\rm nt}_{\rm gen}$, or $P_{\rm gen}^i$. Entropy was estimated as $-\<\log_2 P^i\>$ over the respective generated sample. For the post-selection ensemble ($P_{\rm post}^i$), the distributions were weighted by $Q^i$ computed by SONIA, and the entropy was calculated as $-\<Q(\sigma)\log_2 [P_{\rm gen}(\sigma)Q(\sigma)]\>$ over the generated amino acid sequences.

\subsection{Inference and Probability computation}
Overall workflow is summarized in Fig.~\ref{cartoon}B. VDJ generation models were all inferred using IGoR~\cite{Marcou2018}. Amino acid $P_{\rm gen}$ distributions were all computed using OLGA~\cite{Sethna2019} according to the specified IGoR model parameters. All generated sequences were drawn from the corresponding VDJ generation model using OLGA. Lastly, selection models were all inferred, and evaluated using SONIA. The code for all processes is available on GitHub:\\
IGoR: \url{https://github.com/qmarcou/IGoR}\\
OLGA: \url{https://github.com/zsethna/OLGA}\\
SONIA: \url{https://github.com/statbiophys/SONIA}\\

\subsection{Quantifying variability}
To produce the variances and covariances of Table~\ref{tab:intraindiv} we took the productive data sequences from each individual along with an equivalent number of synthetic sequences drawn from the individual's VDJ generation model. For each sequence we computed $P_{\rm gen}^{\rm univ}$, $Q^{\rm univ}$, and $P_{\rm post}^{\rm univ}$ using the consensus models. The variance and covariance of each quantity was computed over both the data sequences and generated sequences for each individual. These variances and covariances were then averaged over the individual cohort to yield the numbers in Table~\ref{tab:intraindiv}. Error bars are the standard deviation over the cohort.

For Table~\ref{tab:interindiv}, we learned a consensus VDJ generation model $P_{\rm gen}^{\rm univ}$ from nonproductive sequences sampled randomly from all individuals. 300,000 productive sequences were drawn from $P_{\rm gen}^{\rm univ}$ to serve as a generated sequence pool. For data sequences we used 326,000 productive sequences sampled randomly from all individuals. We calculated for each sequence $\sigma$ the individual specific $P_{\rm gen}^i$, $Q^i$, and $P_{\rm post}^i$ for each individual, then calculated the variances and covariances over $i$. Finally we averaged the results over the sequences $\sigma$ from each pool.

The Jensen-Shannon divergence between two distribution $P_1$ and $P_2$ is defined as:
\begin{equation}
  \mathrm{JSD}(P_1,P_2)=\frac{1}{2}\sum_\sigma \left[ P_1(\sigma)\ln\frac{P_1(\sigma)}{\bar P(\sigma)}+P_2(\sigma)\ln\frac{P_2(\sigma)}{\bar P(\sigma)}\right],
\end{equation}
with $\bar P=(P_1+P_2)/2$.

{\bf Acknowledgements.} The work of TM and AMW was supported in
part by grant ERCCOG n. 724208.  The authors have no conflicts of interest.
\medskip

\bibliographystyle{pnas}

\appendix
\setcounter{table}{0}
\renewcommand{\thetable}{S\arabic{table}}%
\setcounter{figure}{0}
\renewcommand{\thefigure}{S\arabic{figure}}%

\begin{figure*}
\begin{center}
\includegraphics[width=0.9\linewidth]{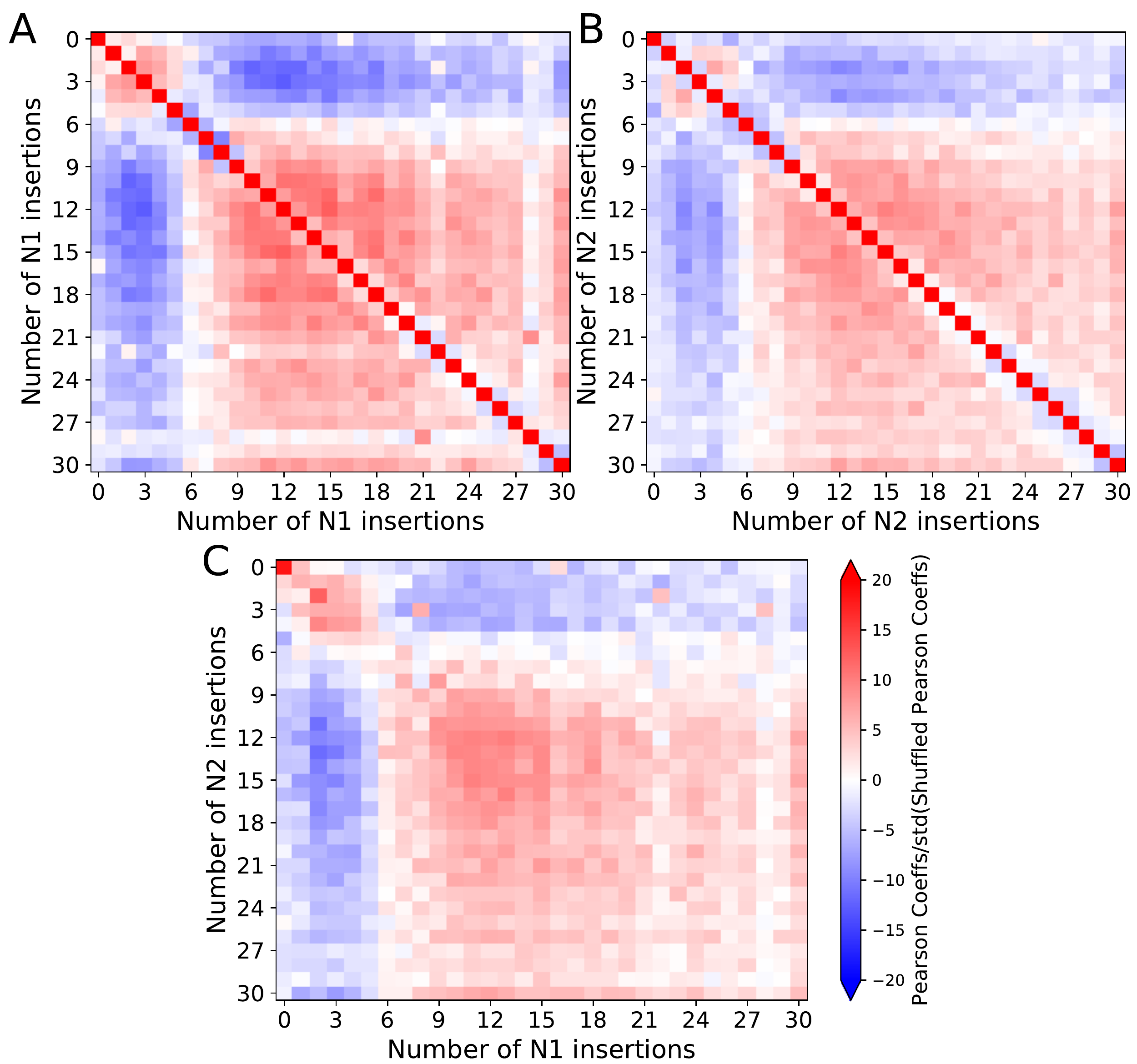}
\caption{Rescaled Pearson coefficients for length insertion distributions. (A) N1-N1 correlations. (B) N2-N2 correlations. (C) N1-N2 correlations. The N1 and N2 distributions are highly correlated over the 651 individual cohort. Rescaling is done by normalizing by the standard deviation of correlation coefficients obtained by shuffling individuals for the two features independently.}
\label{N1N2_corr}
\end{center}
\end{figure*}

\begin{figure*}
\begin{center}
\includegraphics[width=0.9\linewidth]{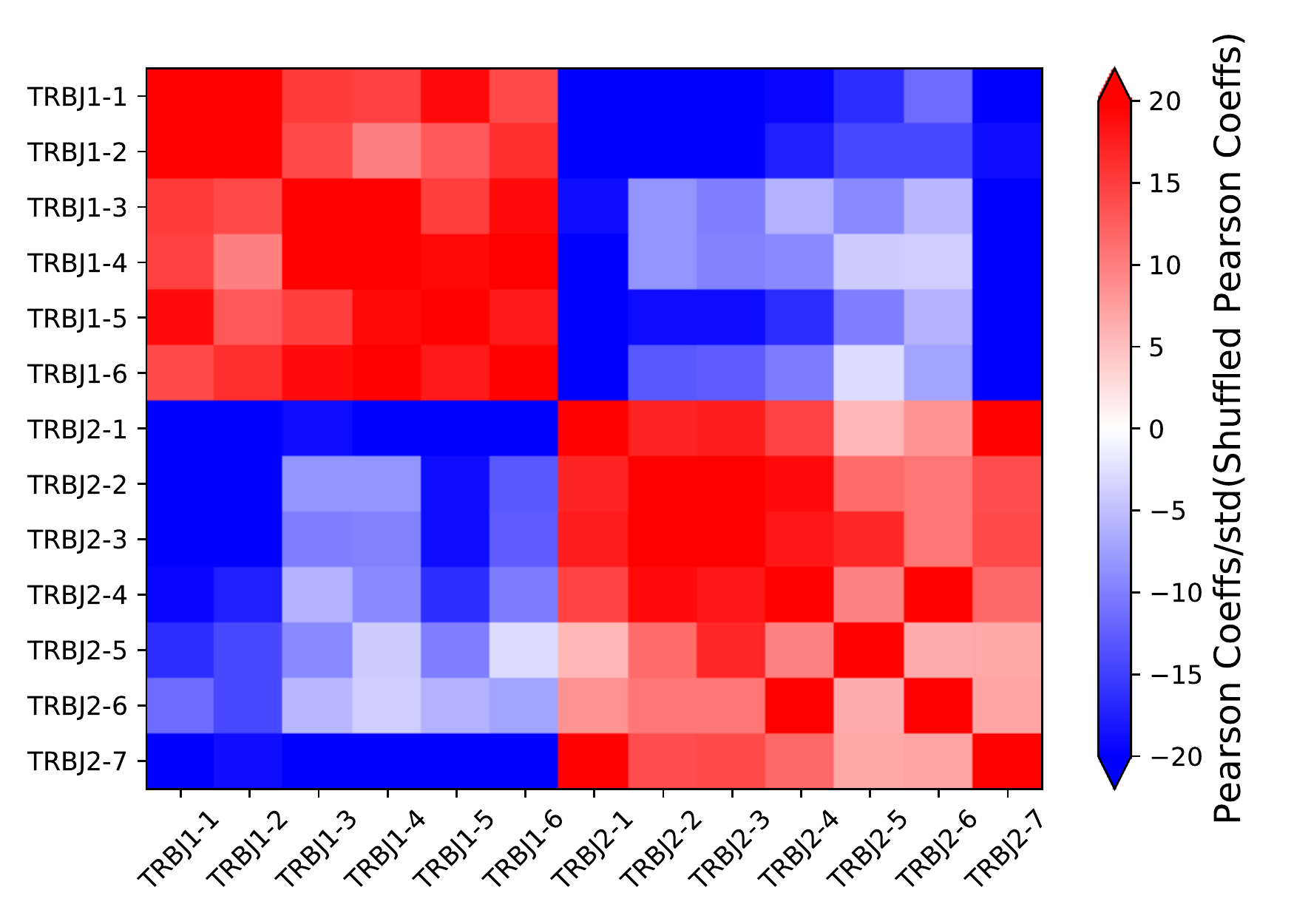}
\caption{Rescaled Pearson coefficients for J-J correlations across the 651 individual cohort. The dominant signal comes from correlations derived from the arrangement of the D and J genes on the chromosome. As genes of the J1 family cannot recombine with the D2 gene, variations in the D usages result in an overall shift in the J1 and J2 gene family usages. This accounts for the strong positive correlation within each J gene family and strong negative correlation between the J1 and J2 families. Rescaling as in S1.}
\label{JJ_corr}
\end{center}
\end{figure*}

\begin{figure*}
\begin{center}
\includegraphics[width=0.9\linewidth]{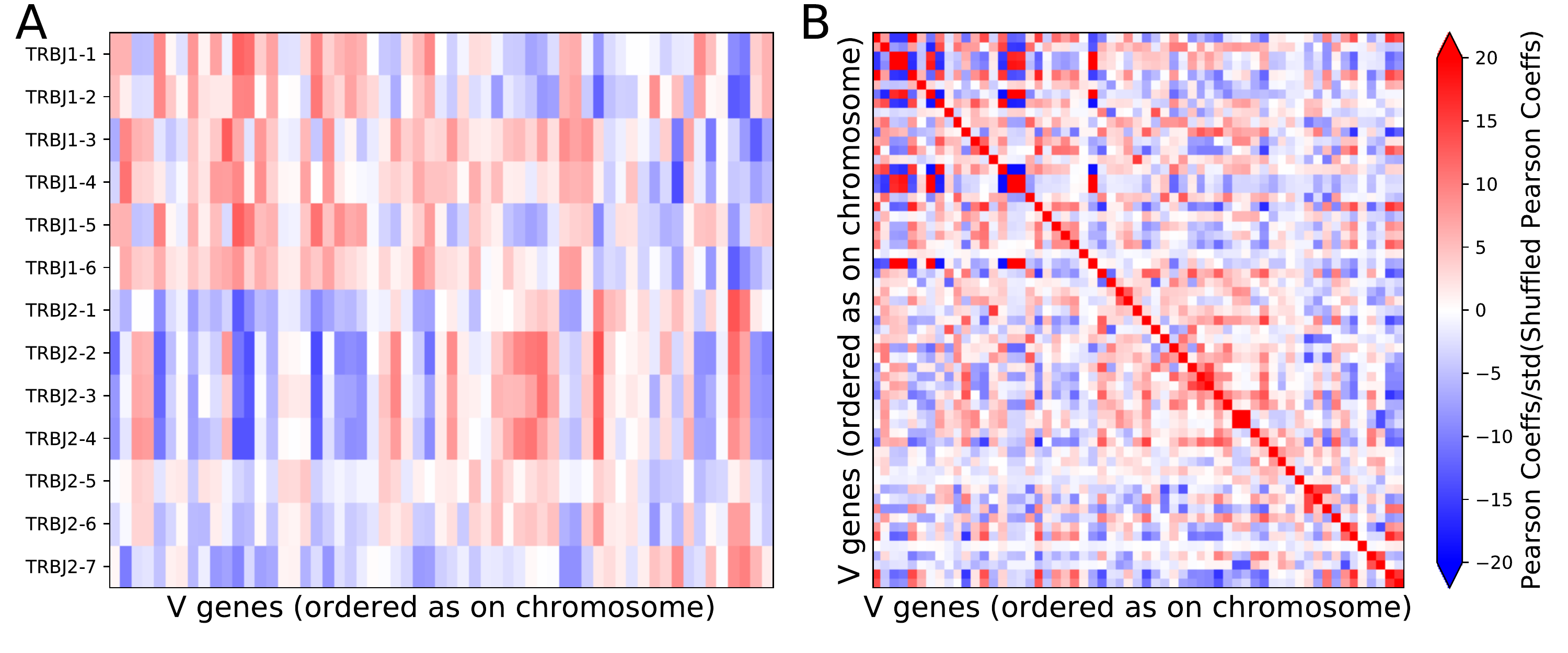}
\caption{(A) Rescaled Pearson coefficients for V-J correlations across the 651 individual cohort. (B) Rescaled Pearson coefficients for V-V correlations. V genes are ordered by position on the chromosome. While large V-J and V-V correlations exist, no obvious chromosomal structure emerges. Rescaling as in S1.}
\label{VJ_corr}
\end{center}
\end{figure*}

\begin{figure*}
\begin{center}
\includegraphics[width=0.7\linewidth]{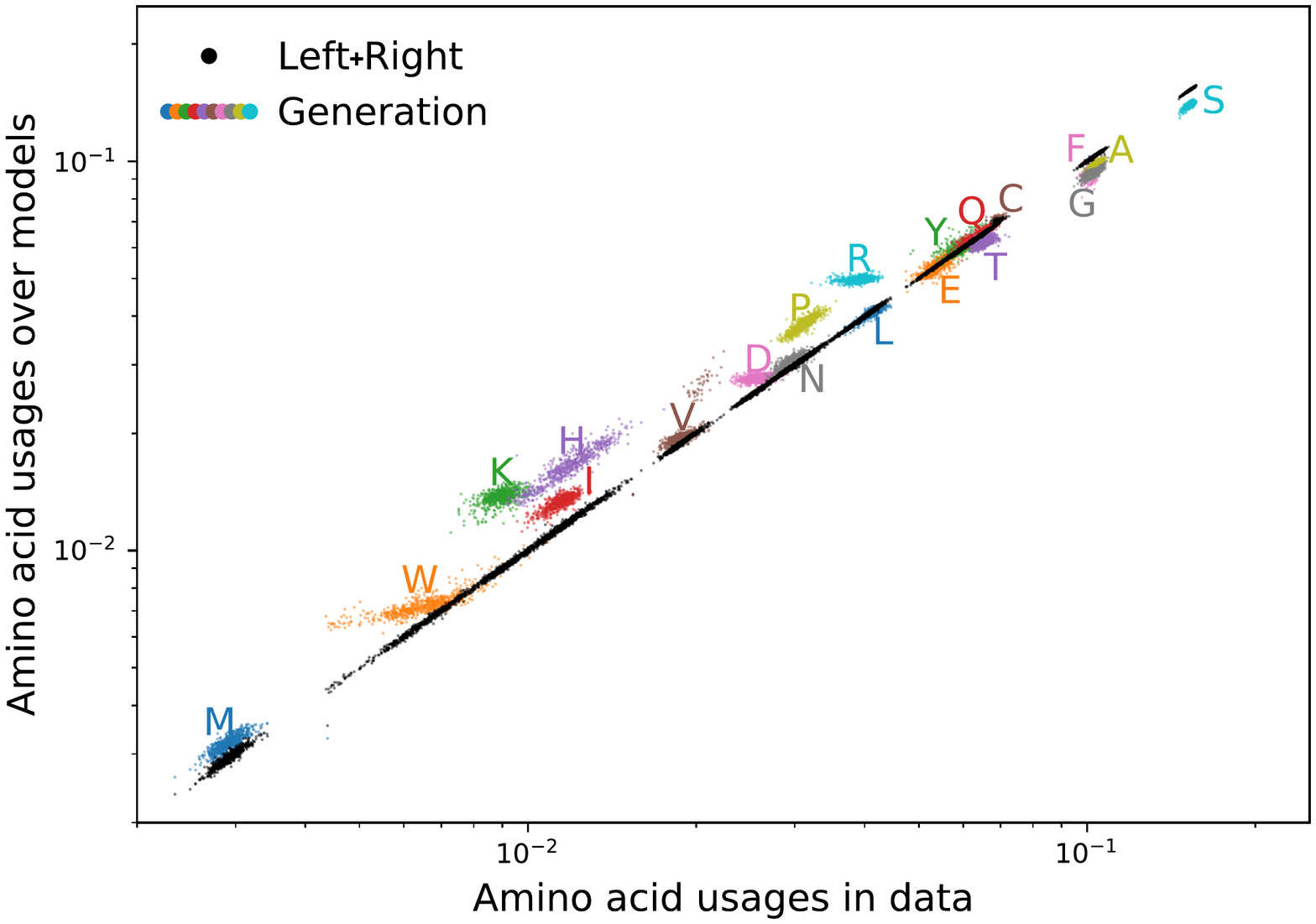}
\caption{Overall amino acid usage in the CDR3. The x-axis is the amino acid usage over the data sequences from a given individual. The y-axis is the amino acid usage over sequences generated from the same individual's VDJ generation model $P_{\rm gen}^i$ (colored dots, each point is an individual), or the same sequences weighted by the $Q^i$ factors from the individual's \textit{Left+Right} selection model (black dots).}
\label{aa_marginals}
\end{center}
\end{figure*}

\begin{figure*}
\begin{center}
\includegraphics[width=0.9\linewidth]{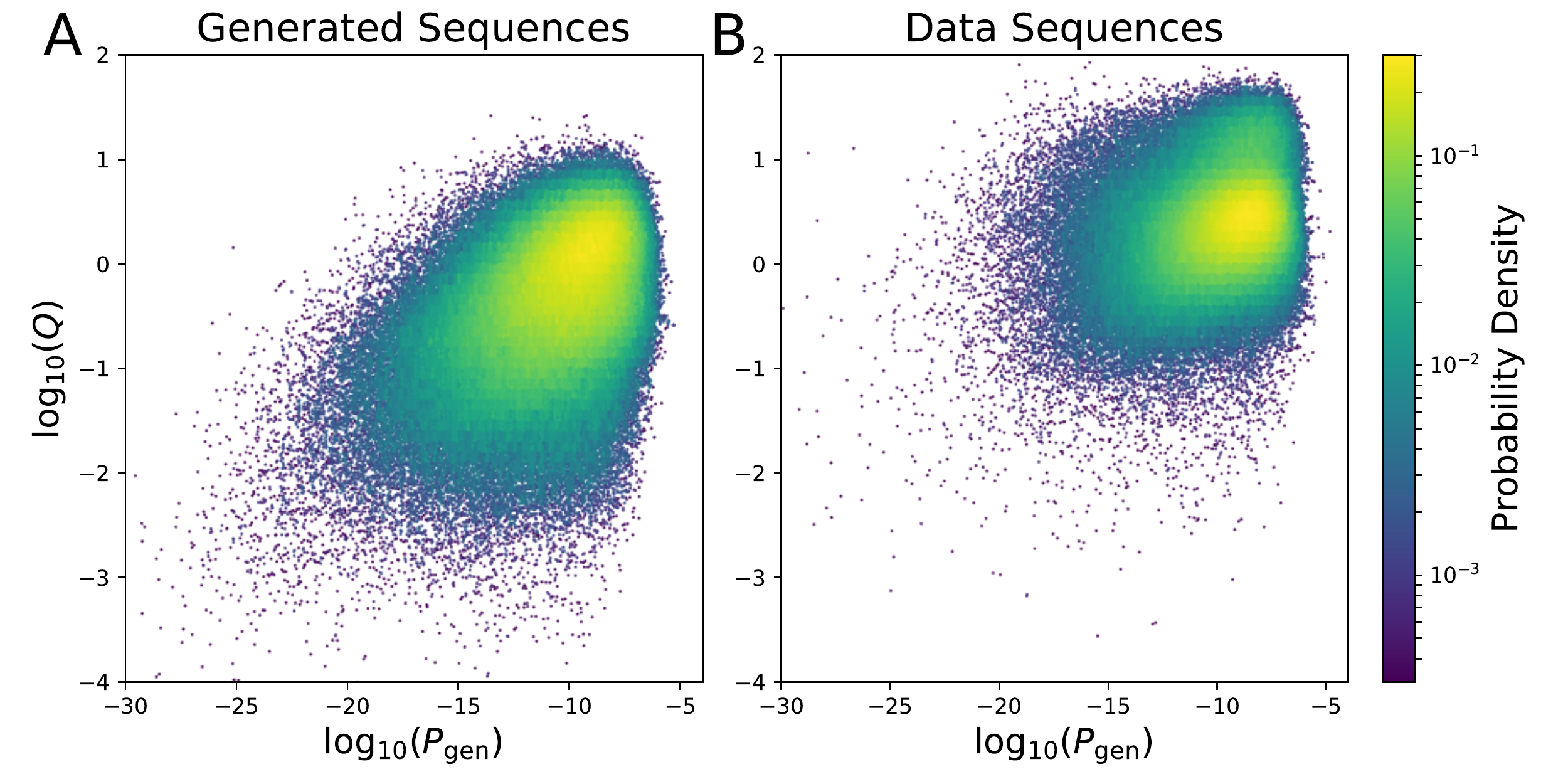}
\caption{Scatter plots of $\log_{10}(Q^{\rm univ})$ vs $\log_{10}(P_{\rm gen}^{\rm univ})$ for (A) generated sequences drawn from $P_{\rm gen}^{\rm univ}$ and (B) data sequences used to infer $\log_{10}(Q^{\rm univ})$. The color scale indicates the local probability density of the points (on a log scale). This visualizes the correlation of $P_{\rm gen}$ and $Q$ as described in Tab.~I. $Q^{\rm univ}$ and $P_{\rm post}^{\rm univ}$ are `universal' models learned from sequences randomly drawn from all individuals. }
\label{pgen_vs_q_scatter}
\end{center}
\end{figure*}

\begin{figure*}
\begin{center}
\includegraphics[width=0.8\linewidth]{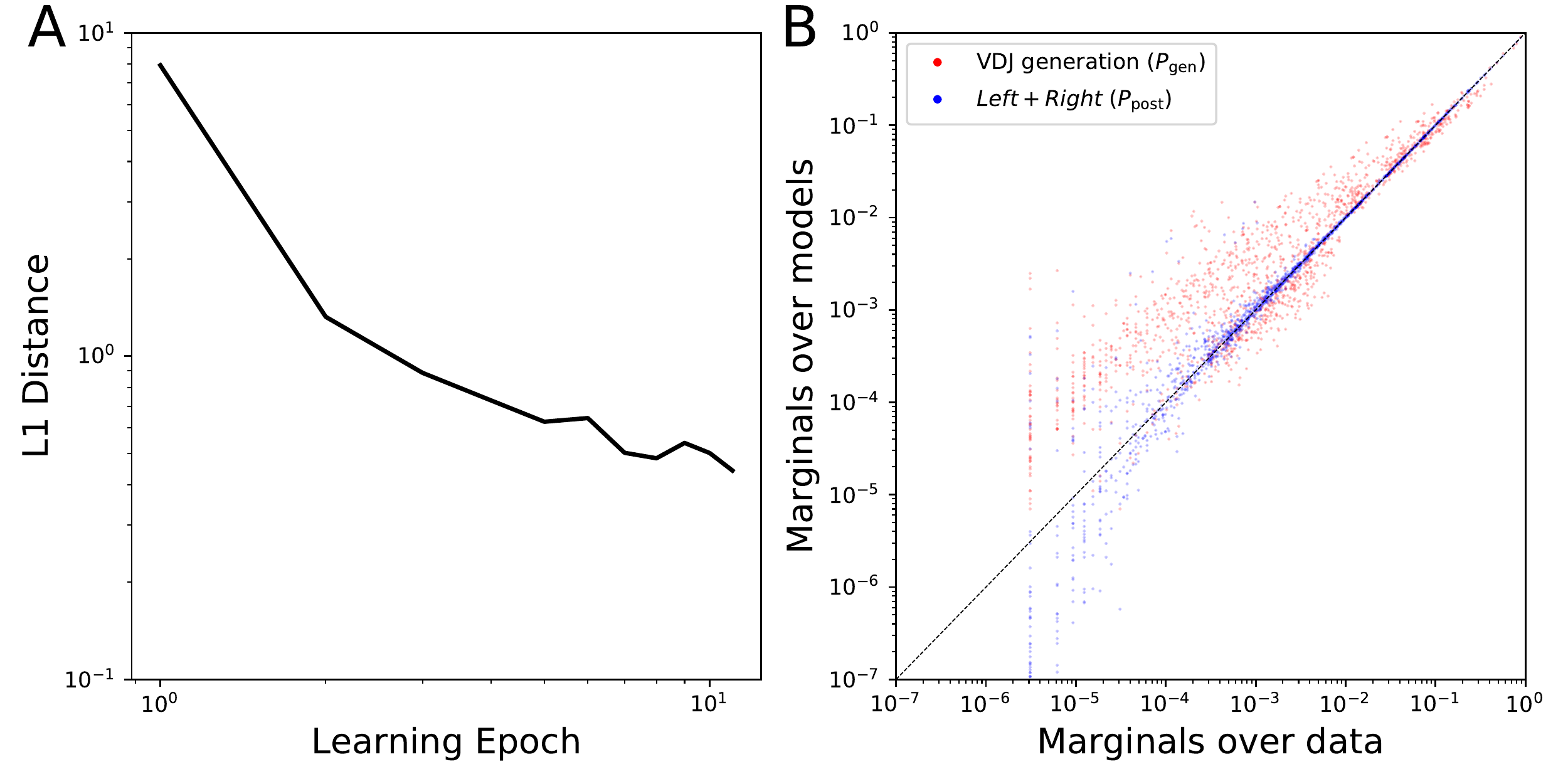}
\caption{Convergence of the universal \textit{Left+Right} model $Q^{\rm univ}$. (A) L1 convergence, per learning epoch, of the marginals (or frequencies) between the data features and the model features. (B) Scatter plot of the feature marginals. The x-axis shows the frequencies of features of the data, while the y-axis show the model prediction for the generation model (red) and for Q-weighted \textit{Left+Right} model (blue). The L1 distance in (A) measures the mean distance between the blue dots and the diagonal.}
\label{converge_fig}
\end{center}
\end{figure*}

\end{document}